**Composition Susceptibility and the Role of One, Two and Three Body Interactions in Glass Forming Alloys: $Cu_{50}Zr_{50}$ vs $Ni_{50}Al_{50}$.**

Chunguang Tang

*School of Materials Science and Engineering,*

*University of New South Wales, Sydney, NSW 2052, Australia*

and Peter Harrowell

*School of Chemistry, University of Sydney, Sydney, NSW 2006 Australia*

Abstract

In this paper we compare the composition fluctuations and interaction potentials of a good metallic glass former, $Cu_{50}Zr_{50}$, and a poor glass former, $Ni_{50}Al_{50}$. The Bhatia-Thornton correlations functions are calculated. Inspired by the observation of chemical ordering at the NiAl surface, we derive a new property, $\hat{R}_{cn}(q)$, corresponding to linear susceptibility of concentration to a perturbation in density. We present a direct comparison of the potentials for the two model alloys, using a $2^{nd}$ order density expansion, establish that the one body energy plays a crucial role in stabilizing the crystal relative to the liquid in both alloys but that the three body contribution to the heat of fusion is significantly larger in NiAl that CuZr.

# 1. Introduction

The rate of crystallization establishes the lifetime of the metastable supercooled liquid and the minimum cooling rate necessary to form a glass. Understanding the factors that determine this rate has been a subject of active research for over 60 years [1]. Classical nucleation theory attributes the rate of crystal nucleation to the influence of three distinct factors: the enthalpy difference between crystal and liquid at the melting point, the dynamics in the liquid state and the effective crystal-liquid interfacial free energy [2]. Glass forming ability, from the perspective of the classical nucleation theory, is the net consequence of the specific values of these disparate quantities in a given liquid. In search of a more unified explanation, researchers have proposed that the kinetic stability of a supercooled liquid to crystallization is determined by local liquid structures that are both stable enough to occur with high frequency but incompatible with a crystal structure [3].

Since the initial suggestion along these lines by Frank in 1952 [4], there has grown a considerable literature on the relation between liquid structure and glass forming ability, largely based on simulation studies of liquid alloys [5,6]. Establishing an explicit connection between liquid structure and crystallization rate has, however, proved a problem. Simply demonstrating the presence of a structure in a liquid that failed to crystallize is not the same as establishing that the structure as responsible for the sluggish ordering kinetics. To establish the role of liquid structure, some sort of direct control of this structure is required. Taffs and Royall [7], for example, have reported on the crystallization of a hard sphere liquid subjected to a bias that favours 5-fold common neighbour coordination. They established a clear dependence of the reduced crystallization time on the magnitude of the bias field, confirming that the liquid structure does indeed influence the rate of crystallization. Perturbations of the liquid structure can result in changes to the crystal structure, complicating the argument. Lee et al [8] studied the kinetics of crystal nucleation in a Ti-Zr-Ni alloy in which the degree of local icosahedral order could be varied with composition. Here the increase in liquid icosahedral order coincided with stabilization of a quasicrystal whose nucleation rate was significantly greater than that of the cubic crystal, effectively concealing whatever influence the liquid structure had on the crystallization of the latter crystal. The relation between competing stable structures and crystallization has been explicitly established in a lattice model of a liquid [9] where it was shown that it was the multiplicity of the stable local structures, rather than the stability of any one particular structure, that was responsible for the stability of the liquid with respect to crystallization.

In a liquid alloy, crystallization typically requires chemical ordering alongside geometrical ordering. Desré and coworkers [10] argued for the kinetic significance of a two-step crystal nucleation in which the first step is a concentration fluctuation to produce a cluster of the same concentration as the crystal followed by crystal nucleation within the droplet. Kelton [11] has emphasised the kinetic importance of long range diffusion in the nucleation kinetics involving composition change. In the case of the equimolar alloys of NiAl and CuZr, the first-forming crystal phase is the equimolar B2 structure, so no segregation is required. Zhang et al [12] have identified two distinct size ratio ranges in ternary mixtures of hard spheres that result in glass formation in compression. In one, all three diameters lie with 10% of each other and so there is little packing benefit to segregation. In the second scenario, the ratio of the largest to the smallest is large, exceeding that needed to drive segregation, and the third particle has a size intermediate between the other two. Compositional fluctuations of some form in the liquid, therefore, represent a feature of liquid structure whose explicit association with the crystallization process promises a useful route to accounting for the variation in crystallization kinetics among different alloys. Comparative studies of similar alloys with very different crystallization rates represent an elegant way of identifying the factors that determine glass forming abilities. In a detailed comparison of the structure in alloys of NiZr and CuZr [13], the former (a non-glass former) was found to exhibit stronger spatial correlations in concentration fluctuations [14] concentration fluctuations than the latter glass-forming alloy. In this case, the composition fluctuations represented a considerably more striking difference than that associated with the frequencies of particular local geometries. Recently [15], we have shown that the anomalously slow crystal growth rate of a glass forming alloy, CuZr, relative to that of a similar (but not glass forming) alloy NiAl, was associated with the difference in composition fluctuations of the two liquids at the crystal-liquid interface. While NiAl exhibits an interfacial structure of alternating Ni and Al layers that extends into the liquid, the slower crystallizing CuZr exhibits almost no chemical ordering at the crystal interface. These results suggest that composition fluctuations in the liquid might provide exactly the structural signature of the glass forming ability of metal alloys.

In this paper, we address the question - can the composition fluctuations observed in the supercooled liquid be correlated with the observed glass forming ability using simulations of $Ni_{50}Al_{50}$ and $Cu_{50}Zr_{50}$? The model and algorithm details are provided in the following Section. In Section 3, we present data for the partial pair correlation functions and the analogous Bhatia-Thornton correlation functions. In Section 4 we look at the linear response of the composition to spatially varying perturbations. In Section 5 with present a detailed examination of the problem of directly comparing the many body potentials for two different alloys

## 2. Models and Algorithms

In this paper we have modelled the alloys using many bodied Embedded Atom Model (EAM) potentials of CuZr due to Mendelev et al [16] and NiAl potential due to Mishin et al [17]. We provide a detailed discussion of the mathematical form of these potentials in Section 5. The simulations were carried out under constant temperature and pressure conditions using the LAMMPS algorithms [18]. The pressure was set to zero and 2744 atoms of each species were used. The relevant liquid and crystal phases were equilibrated at specified temperatures for analysis. More details can be found in ref. [19].

## 3. The Spatial Correlation of Composition Fluctuations in Liquid $Cu_{50}Zr_{50}$ and $Ni_{50}Al_{50}$

The partial pair radial distribution functions represent the standard characterization of structure in a liquid mixture. These functions for the supercooled CuZr and NiAl are plotted in Fig. 1 using the respective lattice spacings $a_o$ as a unit of length. The respective temperatures were chosen to produce roughly the same chemical potential difference relative to the crystal state. We find that a) the first peak in Ni-Al is higher than that of Cu-Zr, indicating a greater degree of compositional ordering in the former, and b) the nearest neighbour distances Ni-Ni and Ni-Al deviate from the simple additivity exhibited in the CuZr mixture. (Specifically, the Ni-Al distance is shorter than that obtained by averaging the Ni-Ni and Al-Al lengths.)

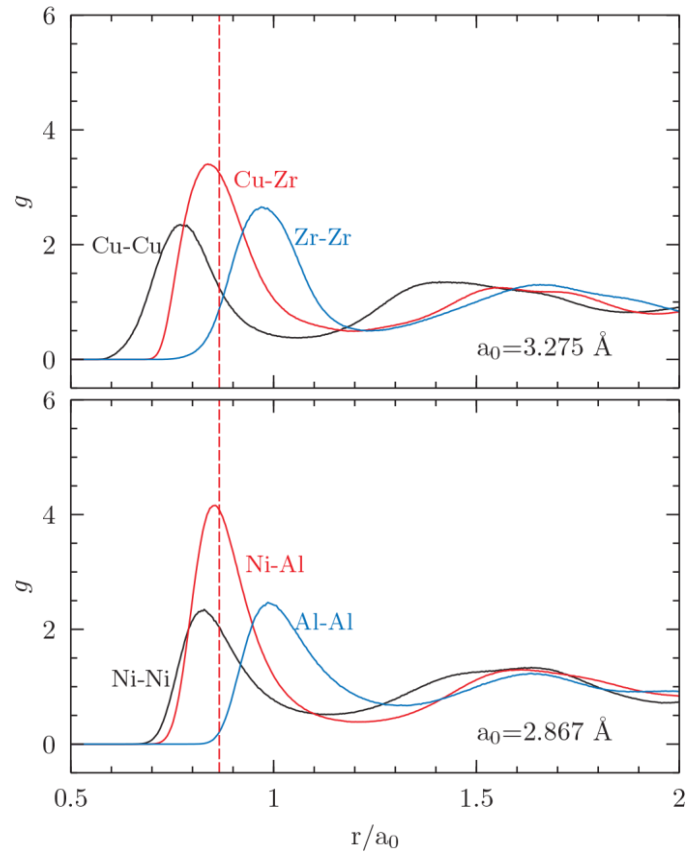

**Figure. 1.** The partial radial distribution functions for liquid $Cu_{50}Zr_{50}$ at T = 1100K (upper panel) and liquid $Ni_{50}Al_{50}$ at T = 1300K (lower panel). The distances are scaled by the respective crystal lattice spacing $a_o$ as indicated. The vertical red dashed line corresponds to $r/a_o = \sqrt{3}/2$. The temperatures were chosen so that both liquids are at roughly the same degree of supercooling as measured by the chemical potential difference between liquid and crystal.

To put these observations regarding the liquid structure into some sort of context, it is useful to compare the distribution of atomic separations in the liquid and crystal state. In Fig. 2 we plot the partial radial distribution functions for the crystal for each alloy. The geometric constraints of the B2 crystal (a body centred structure with different species in the centre and corner positions), impose stringent conditions on the relative nearest neighbour distances, namely, that a) the lattice spacing is set by the neighbour distance between the larger species, b) the nearest neighbour separation between like species are equal and c) the distance between neighbouring unlike species is determined completely by the large particle size with the first peak occurring at $r/a_o = \sqrt{3}/2$. All three of these correlations are evident in the distributions plotted in Fig. 2. The crystal order is associated with a large number of unlike species neighbours, as seen by the heights of the Ni-Al and Cu-Zr peaks. An essential characteristic of the crystal structure is that all lengths are set by the nearest neighbour separation between the larger particles. This means, for example, that the nearest neighbour distance between the smaller species, Ni and Cu, respectively, increases significantly in going from the liquid to the crystal. Looking back at the liquid pair correlation functions in Fig. 1 we note that the position of the Ni-Al peak occurs at almost exactly the crystal distance, i.e. $r/a_o = \sqrt{3}/2$, while the first maximum in the Cu-Zr peak occurs at a shorter distance. This result suggests that the packing in the NiAl liquid reflects some degree of local crystal-like organization while that in CuZr, less so.

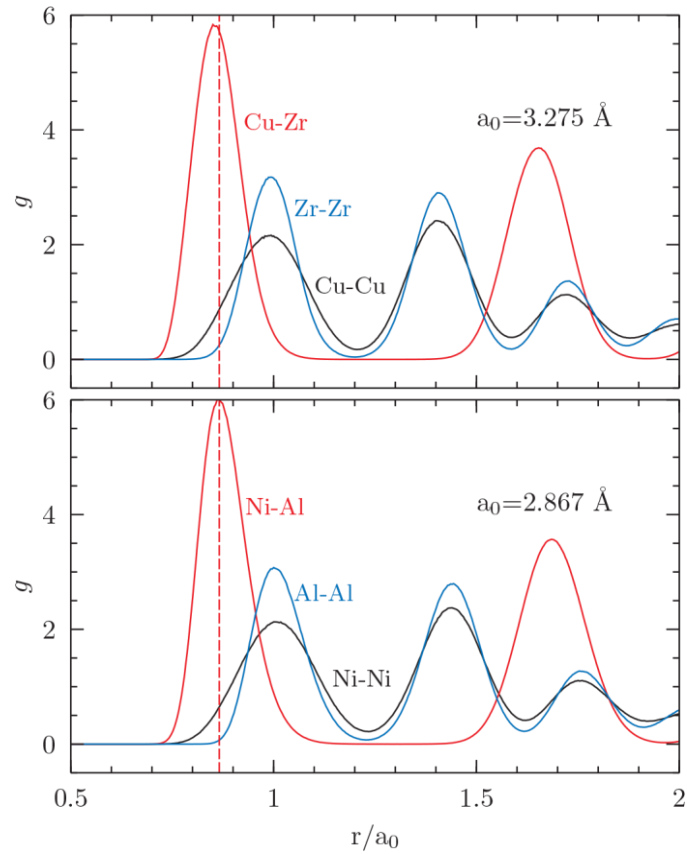

**Figure 2.** The particle radial distribution functions of the B2 crystal phases for CuZr at T = 1100K (upper panel) and NiAl at T = 1300K (lower panel). The vertical red dashed line corresponds to $r/a_o = \sqrt{3}/2$, the distance between unlike species in the B2 structure. Note that the distance is scaled by the relevant lattice constant $a_0$ in each case.

The deviation of the composition of the first coordination shell from that of the bulk concentration represents the local deviation from random mixing in the liquid. The Warren-Crowley parameter [20] $\alpha_{ab} = 1 - \frac{f_{ab}}{c_b}$ is defined in terms of $f_{ab}$, the average fraction of the first neighbours of an $a$ particle that are species $b$, and $c_b$, the bulk value of the number concentration of species b, i.e. $c_b = \frac{N_b}{N_a + N_b}$. Note that random mixing in the local coordination shell will give a Warren-Crowley parameter of zero, while a tendency for unlike species to aggregate (i.e. chemical ordering) will result in a negative value of $\alpha_{ab}$. The values of the Warren-Crowley parameters for our two alloys are $\alpha_{CuZr}$ = -0.32, $\alpha_{ZrCu}$ = -0.0042, $\alpha_{NiAl}$ = -0.26 and $\alpha_{AlNi}$ = 0.029 based on the liquid states of CuZr at 1100 K and NiAl at 1300 K, respectively. The small magnitudes of $\alpha$ for the neighbours of the larger atoms, Zr and Al, indicate that their larger coordination shells correspond more closely to the bulk composition. The smaller atoms, however, show a clear preference to associate with the larger atoms, with the tendency to chemically order being larger in Cu than Ni.

If the nature of compositional ordering distinguishes the two alloys, then an alternative representation of correlations that specifically addresses composition fluctuations might be useful. Such an alternative to the particle distribution functions was proposed by Bhatia and Thornton [14] who pointed out that the species specific structure factors could be transformed, by linear combination, to a new set of correlation functions related to the density-density (nn), concentration-concentration (cc) and density-concentration (nc) fluctuations. Here $c = c_1$, the number concentration of species 1. The measurement of Bhatia-Thornton (BT) structure factors using neutron scattering has been reviewed by Fischer et al [21]. Salmon and co-workers [22-24] have demonstrated the utility of the BT structure factors in determining the intermediate structure in network glasses. The BT radial distribution functions $g_{nn}(r)$, $g_{cc}(r)$ and $g_{nc}(r)$ can be obtained from the partial correlation functions $g_{11}(r)$, $g_{22}(r)$ and $g_{12}(r)$ via the linear relations [21],

$$g_{nn}(r) = c_1^2 g_{11}(r) + c_2^2 g_{22}(r) + 2 c_1 c_2 g_{12}(r) \tag{1}$$

$$g_{cc}(r) = c_1 c_2 [g_{11}(r) + g_{22}(r) - 2 g_{12}(r)] \tag{2}$$

$$g_{nc}(r) = c_1[g_{11}(r) - g_{12}(r)] - c_2[g_{22}(r) - g_{12}(r)] \tag{3}$$

where $c_1$ and $c_2$ are the number fractions of species 1 and 2.

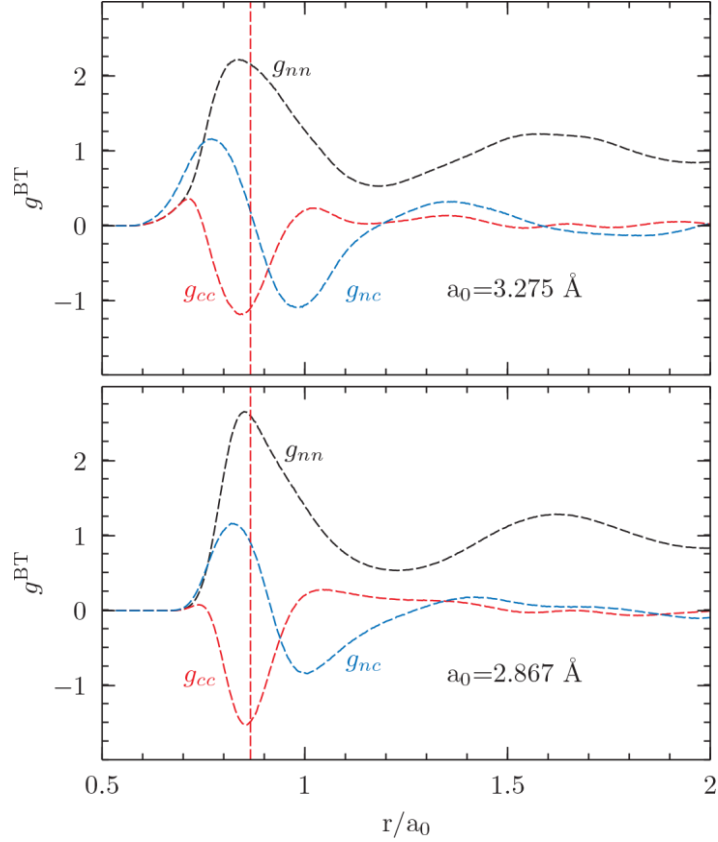

**Figure 3.** The three Bhatia-Thornton radial distribution functions $g_{nn}(r)$, $g_{cc}(r)$ and $g_{nc}(r)$ for liquid $Cu_{50}Zr_{50}$ (upper panel) and $Ni_{50}Al_{50}$ (lower panel) at T = 1100K and 1300K, respectively.

In Fig. 3 we have plotted the BT radial distribution functions for CuZr and NiAl. We find that the first peak in the $g_{nn}(r)$ occurs at roughly the position of the peaks in $g_{NiAl}$ and $g_{CuZr}$, respectively. The position of this peak coincides, in both mixtures, with the position of the maximum *anti-correlation* in concentration fluctuations, although the width of the $g_{nn}$ peak is considerably broader than the associated minimum in $g_{cc}$.) These results indicate that density correlations are coupled to the local chemical order. Mathematically, these results follow from the dominance of the $g_{12}$ peak in both alloys. The cross-coupling $g_{nc}$ switches from positive to negative with increases separation, within the span of the density-density correlation. This cross-coupling reflects the fact that the density correlations combine both 11 and 12 correlation pairs.

Previously, Kaban et al [13] reported the analysis of scattering data indicated that the BT structure factor $S_{cc}(q)$ (obtained through the Fourier transform of $g_{cc}(r)$) for $Cu_{65}Zr_{35}$ exhibited significantly smaller amplitude than the analogous quantity for $Ni_{64}Zr_{36}$, the latter being a poorer glass former. In Fig. 4 we have plotted the BT structure factors for the two alloys. We find little difference between the amplitudes of $S_{cc}(q)$ in the two alloys studied in this paper. In the CuZr alloy, the peaks in $S_{cc}(q)$ do appear to have shoulders that are absent in NiAl, suggesting multiple length scales associated with the compositional correlation.

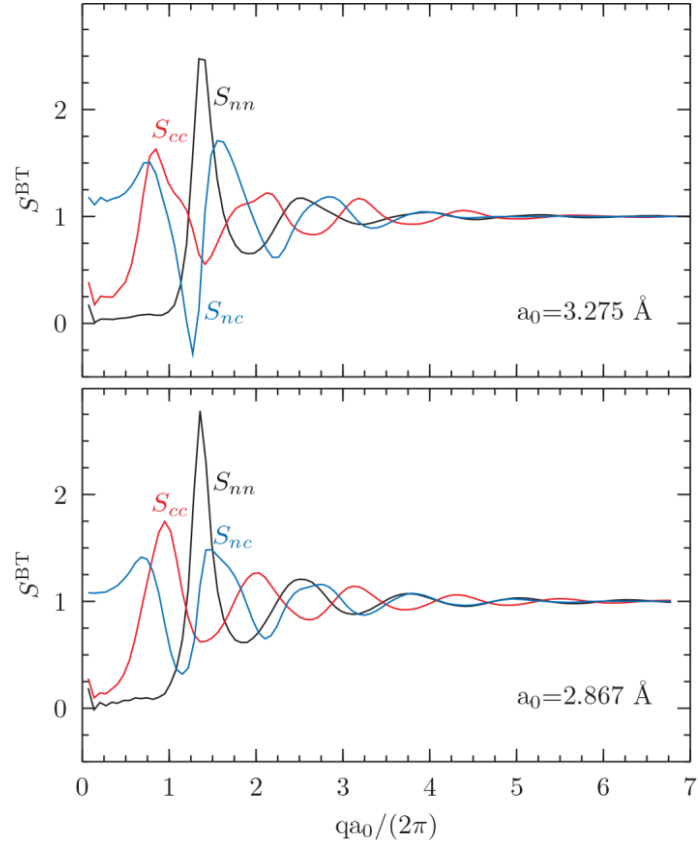

**Figure 4.** The three Bhatia-Thornton structure factors $S_{nn}(r)$, $S_{cc}(r)$ and $S_{nc}(r)$ for $Cu_{50}Zr_{50}$ (upper panel) and $Ni_{50}Al_{50}$ (lower panel) at T = 1100K and 1300K, respectively.

In the context of these specific calculations, we conclude that the BT correlations do not provide any obvious better differentiation of the structure of the two liquid alloys than that of the partial pair correlations functions.

**4. The Susceptibility of Composition Fluctuations to Density Variations in a Liquid Mixture: Linear Response Theory.**

Along with information about the distribution of length scales in a liquid, the pair correlation functions also express the susceptibility, in the linear response limit, of the density with respect to an applied field. This perspective suggests a new approach to connecting liquid structure and crystallization kinetics. Previously [15], we have noted that our two alloys were differentiated by the degree of compositional ordering in the liquid adjacent to the crystal interface. This suggests that it is the susceptibility of the liquid to the perturbation provided by the interface that might most clear identify glass forming ability.

First, we should note the rigorous connection between the Bhatia-Thornton structure factors and thermodynamic properties of the binary mixture [14], namely

$$S_{nn}(0) = n_o k_B T \kappa_T + \delta^2 S_{cc}(0)$$
$$S_{cc}(0) = k_B / \left(\partial^2 G / \partial c_A^2\right)_{T,p,N} \quad (4)$$
$$S_{nc}(0) = -\delta S_{cc}(0)$$

where G is the Gibbs free energy per particle, $\kappa_T$ is the isothermal compressibility, $n_o$ is the atomic number density and the dilation factor $\delta = \frac{N}{V}(v_1 - v_2)$ where $v_1$ and $v_2$ are the particle molar volumes per particle of the two species.

A direct perturbation would be to apply a field u(r) and measure the response in terms of the density [25]. We could imagine two types of potentials: $u_c(r)$ that couples directly to concentration variations and $u_n(r)$ that couples to the total density. The linear responses to these potentials in terms of the density and concentration fields are:

$$\delta c(\vec{r}_1) = \int d\vec{r}_2 G_{cc}(\vec{r}_1, \vec{r}_2) u_c(\vec{r}_2) \quad (7)$$

$$\delta c(\vec{r}_1) = \int d\vec{r}_2 G_{nc}(\vec{r}_1, \vec{r}_2) u_n(\vec{r}_2) \quad (8)$$

$$\delta n(\vec{r}_1) = \int d\vec{r}_2 G_{nn}(\vec{r}_1, \vec{r}_2) u_n(\vec{r}_2) \quad (9)$$

$$\delta n(\vec{r}_1) = \int d\vec{r}_2 G_{nc}(\vec{r}_1, \vec{r}_2) u_c(\vec{r}_2) \quad (10)$$

where the field $u(\vec{r})$ is assumed to be small enough to justify retaining only the linear term and $\delta c(\vec{r}) = c(\vec{r}) - c_o$ and $\delta n(\vec{r}) = n(\vec{r}) - n_o$. The identity of the susceptibilities $G_{\alpha\beta}(r_1, r_2)$ is most simply expressed in the Fourier transformed versions of Eq. 7-10, i.e.

$$\delta \hat{c}(q) = S_{cc}(q) \delta \hat{u}_c(q) \qquad (11)$$

$$\delta \hat{c}(q) = S_{cn}(q) \delta \hat{u}_n(q) \qquad (12)$$

$$\delta \hat{n}(q) = S_{nn}(q) \delta \hat{u}_n(q) \qquad (13)$$

$$\delta \hat{n}(q) = S_{cn}(q) \delta \hat{u}_c(q) \qquad (14)$$

where $\delta \hat{c}(q)$ represents the Fourier transform $\delta c(r)$. The essential result of this analysis is to establish that the structure factors can be regarded as wavevector dependent susceptibilities of the liquids, expressing the magnitude of the concentration or density variation arising from the application of the appropriate spatially varying field.

In the case of an interface, it is the response of the concentration to a density variation δn(r) rather than an external field $u_n$ (r) that we are interested in. A central tenant of density functional theory is that there is a one-to-one relationship between the potential and the density change [25]. This means that we can eliminate the field $u_n$(r) in favour of the density fluctuation δn(r) via a convolution, the desired expression relating a concentration change $\delta \hat{c}(q)$ to a density fluctuation $\delta \hat{n}(q)$. The details are provided in the Appendix. The result is a new response function $R_n(\vec{r}_1, \vec{r}_2)$ whose Fourier transform in the isotropic liquid satisfies

$$\delta \hat{c}(q) = \hat{R}_{cn}(q) \delta \hat{n}(q) \qquad (15)$$

where

$$\hat{R}_{cn}(q) = \frac{S_{nc}(q)}{S_{nn}(q)} \tag{16}$$

Here the response function $\hat{R}_{cn}(q)$ is the magnitude of the concentration fluctuation with wavevector q relative to the magnitude of the density fluctuation at the same wavevector. A large value of $\hat{R}_{cn}(q)$ implies large concentration variations while the disappearance of $\hat{R}_{cn}(q)$ at a given wavevector indicates that there is, up to linear response, no concentration change associated with the density variation at that q.

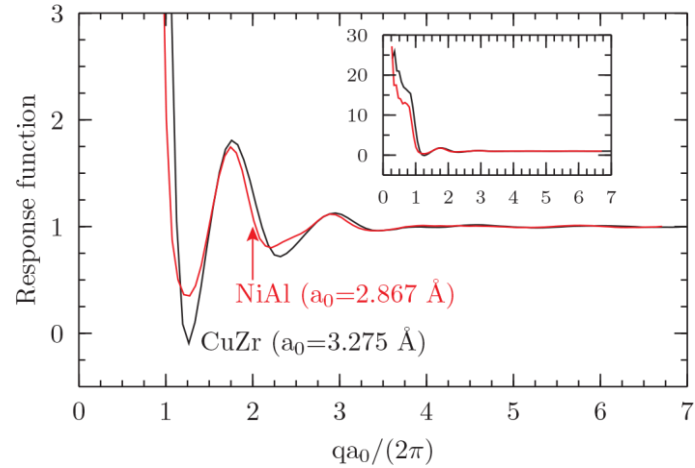

**Figure 5**. The response function $\hat{R}_{cn}(q)$, as defined in Eq. 16, corresponding to the q-dependent response of the composition to a density variation. Below $qa_o/(2\pi) = 1$, the small value of $S_{nn}(q)$ renders the calculation of $\hat{R}_{cn}(q)$ statistically unreliable.

We plot the values of the susceptibility $\hat{R}_{cn}(q)$ for NiAl and CuZr in Fig. 5. The large response function at small q is a consequence of the small density fluctuations at these wavevectors associated with the limited compressibility of the liquid. The smaller the magnitude of the intrinsic fluctuations, the larger the applied field required to generate a given density fluctuation at that q and, hence, the larger the response seen in the composition.

The interesting feature of the plotted response function is the vanishing of $\hat{R}_{cn}(q)$ for the CuZr mixture at a wavevector slightly larger than that associated with crystal ordering – i.e. $qa_o/2\pi = 1$. The vanishing of $\hat{R}_{cn}(q)$ indicates an insensitivity of the concentration to density variations at this wavevector. This is consistent with the absence of compositional ordering at the liquid surface in the $Cu_{50}Zr_{50}$ liquid. The NiAl alloy, in contrast, retains a significant response at the crystal wavevector and exhibits clear compositional ordering at the liquid surface. This observation must be qualified by noting that $\hat{R}_{cn}(q)$ changes dramatically over the range of q associated with crystal order and so it is not clear how significant the vanishing of the response function is.

## 5. On Comparing Interaction Potentials between Different Alloys.

Any difference in crystallization kinetics between two simulated alloys must ultimately originate in the differences in their interaction potentials. When the potentials are of the many-bodied type – such as the Finnis-Sinclair potential [26], the Embedded Atom Model (EAM) of Daw and Baskes [27] and the 'glue' potential of Parrinello and co-workers [28] - a meaningful comparison is a non-trivial task. These potentials have the following general form,

$$V = \frac{1}{2}\sum_{i,j}^{N} \phi(r_{ij}) + \sum_{i=1}^{N} F(\rho_i) \qquad (17)$$

where $F(\rho_i)$ is the many-bodied contribution to the potential energy arising from electron delocalization and is a function of the local density $\rho_i$ about particle i given by

$$\rho_i = \sum_{j}^{N} w(r_{ij}) \qquad (18)$$

where w(r) is a monotonically decaying function, explicitly defined as part of the EAM potential, that determines the extent of the neighbour environment. Note that the magnitude of $\rho_i$ will depend on the amplitude and extent of the function w(r). As the local density is used only as input to the embedding field $F(\rho_i)$, the actual magnitude of $\rho_i$ will be accommodated within the parameters of F. This means that the absolute magnitude of $\rho_i$ in EAM potentials is arbitrary, a point that shall need to be kept in mind when comparing the results for different alloys.

The problem with a potential energy of the form expressed in Eq. 17 is that it is invariant to the following linear transformation [28],

$$\hat{\phi}(r) = \phi(r) + 2\lambda\rho(r)$$
$$\hat{F}(\rho_i) = F(\rho_i) - \lambda\rho_i$$
(19)

The continuous family of potentials $\{\hat{F}, \hat{\phi}\}$, parameterised by $\lambda$, that can be generated by the transformation in Eq. 19 all result in exactly the same energy for any given configuration. The transformation has simply mixed the 2-body and many-body components (as described by the $F(\rho_i)$) such that the changes cancel each other. This means that *some* differences between the potentials for two different mixtures do not correspond to any actual physical difference at all and so are irrelevant in terms of differentiating the behaviour of the two alloys. To properly compare the interactions of different alloys, therefore, we must first find a representation of the potential energy that is independent of the transformation of Eq. 19. Just such a unique representation of the particle interactions can be obtained through an approximation due to Foiles [29], based on the assumption that the local density $\rho_i$ is narrowly distributed about the average $\bar{\rho}$. The following analysis is predicated on the fields $\{F_\alpha\}$ being explicit analytic functions of the respective local densities $\rho_\alpha$. In applying this approach to a mixture, we start with generalizing of Eqs. 17 and 18 for the potential energy of the mixture,

$$V = \sum_{\alpha} \sum_{i} F_{\alpha}(\rho_{\alpha,i}) + \frac{1}{2} \sum_{\alpha,\beta} \sum_{i,j \neq i} \phi_{\alpha\beta}(r_{ij}) \tag{20}$$

with

$$\rho_{\alpha,i} = \sum_{\beta} \sum_{j}^{N_{\beta}} w_{\alpha\beta}(r_{ij}) \tag{21}$$

We shall focus here on binary mixtures. The embedding field term $F(\rho_\alpha)$ for each of the metals is plotted in Fig.6 along with the distribution in local density.

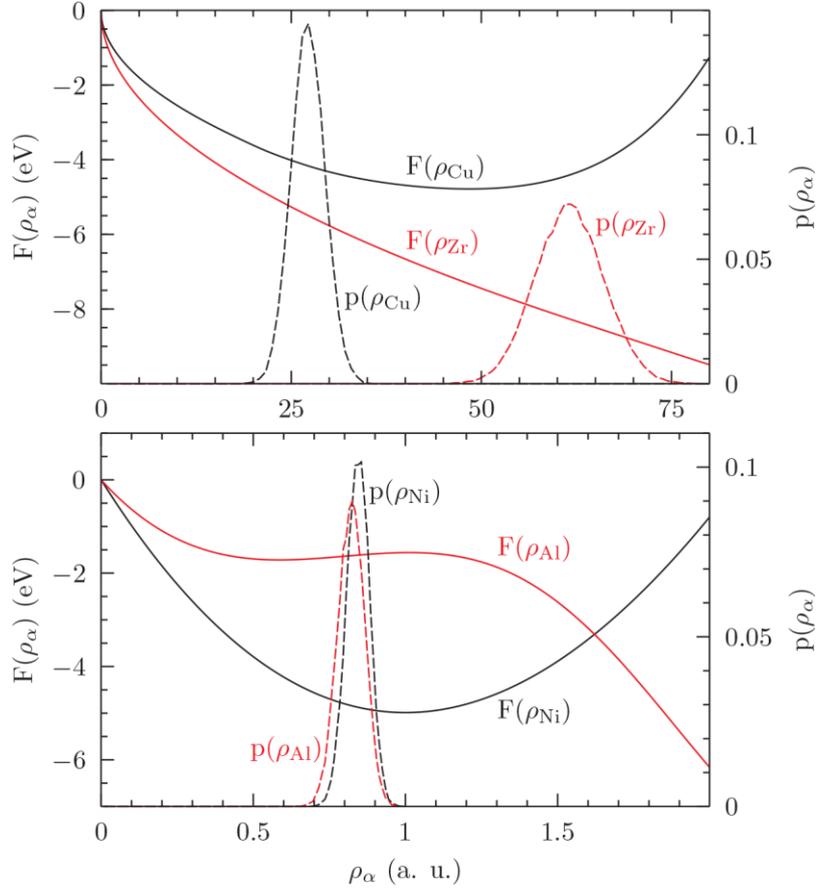

**Figure 6.** The embedding field $F(\rho_\alpha)$, as introduced in Eq. 20, plotted as a function of $\rho_\alpha$ for a) $\alpha$ = Cu and Zr, and b) $\alpha$ = Ni and Al. Also included in each plot is the probability distribution $p(\rho_\alpha)$ of the local density $\rho_\alpha$ in liquid CuZr at 1100 K and liquid NiAl at 1300 K, respectively. Note that the absolute magnitude of the values of the local density $\rho_\alpha$ cannot be meaningfully compared between different alloys as explained in the text.

Given the narrow distributions observed in the local density, as shown in Fig. 6, one route to simplifying the many-body character of the EAM potential, without sacrificing accuracy, is to expand about the average local density. Expanding to second order about the average local densities $\bar{\rho}_\alpha$ we find,

$$V \approx F_o(\bar{\rho}_1, \bar{\rho}_2) + \frac{1}{2}\sum_{i \neq j}\left(\psi_{11}(r_{ij}) + 2\psi_{12}(r_{ij}) + \psi_{22}(r_{ij})\right) + R_3 \tag{22}$$

where the 2-body interactions are

$$\psi_{11}(r) = \phi_{11}(r) + 2[F_1'(\bar{\rho}_1) - \bar{\rho}_1 F_1''(\bar{\rho}_1)]w_{11}(r) + F_1''(\bar{\rho}_1)w_{11}^2(r) \tag{23}$$

$$\psi_{22}(r) = \phi_{22}(r) + 2[F_2'(\bar{\rho}_2) - \bar{\rho}_2 F_2''(\bar{\rho}_2)]w_{22}(r) + F_2''(\bar{\rho}_2)w_{22}^2(r) \tag{24}$$

$$\psi_{12}(r) = \frac{1}{2}(\tilde{\psi}_{12}(r) + \tilde{\psi}_{21}(r)) \tag{25}$$

and where

$$\tilde{\psi}_{12}(r) = \phi_{12}(r) + 2[F_1'(\bar{\rho}_1) - \bar{\rho}_1 F_1''(\bar{\rho}_1)]w_{12}(r) + [F_1''(\bar{\rho}_1)]w_{12}^2(r) \tag{26}$$

$$\tilde{\psi}_{21}(r) = \phi_{21}(r) + 2[F_2'(\bar{\rho}_2) - \bar{\rho}_2 F_2''(\bar{\rho}_2)]w_{21}(r) + [F_2''(\bar{\rho}_2)]w_{21}^2(r) \tag{27}$$

Standard notation for the derivatives $F_a'(\bar{\rho}_a) = \left.\frac{dF_a}{d\rho_a}\right|_{\rho_a=\bar{\rho}_a}$ and $F_a''(\bar{\rho}_a) = \left.\frac{d^2 F_a}{d\rho_a^2}\right|_{\rho_a=\bar{\rho}_a}$ has been used. The uniform field term is

$$F_o = \sum_i (F_1(\bar{\rho}_1) - \bar{\rho}_1 F_1'(\bar{\rho}_1) + \frac{\bar{\rho}_1^2}{2} F_1''(\bar{\rho}_1)) + \sum_i (F_2(\bar{\rho}_2) - \bar{\rho}_2 F_2'(\bar{\rho}_2) + \frac{\bar{\rho}_2^2}{2} F_2''(\bar{\rho}_2)) \quad (28)$$

and the 3-body term, $R_3$, is given by

$$R_3 = \frac{1}{2} F_1'' \left[ \sum_{i \ne j \ne k} w_{11}(r_{ij}) w_{11}(r_{ik}) + \sum_{i \ne j \ne k} w_{12}(r_{ij}) w_{12}(r_{ik}) + \sum_{i \ne j \ne k} w_{11}(r_{ij}) w_{12}(r_{ik}) + \sum_{i \ne j \ne k} w_{12}(r_{ij}) w_{11}(r_{ik}) \right]$$

$$+ \frac{1}{2} F_2'' \left[ \sum_{i \ne j \ne k} w_{22}(r_{ij}) w_{22}(r_{ik}) + \sum_{i \ne j \ne k} w_{21}(r_{ij}) w_{21}(r_{ik}) + \sum_{i \ne j \ne k} w_{22}(r_{ij}) w_{21}(r_{ik}) + \sum_{i \ne j \ne k} w_{21}(r_{ij}) w_{22}(r_{ik}) \right]$$

(29)

Note in Eqs. 22-29, the summations over particles only apply if the particles match the specified types. We shall now apply this general analysis to the specific cases of the CuZr potential of Mendelev et al [16] and the NiAl potential due to Mishin et al [17].

In Fig. 7 we plot the 2-body potentials and in Figs. 8 and 9, the 3-body contributions as calculated using Eqs. 26-29. We find some striking differences between the effective 2-body potentials for the two alloys. In the case of NiAl, we note that i) the three potentials exhibit rough additivity in terms of length scale, ii) the Ni-Ni interaction has a significant attractive well and iii) the Al-Al interaction is only slightly attractive. In contrast, the effective 2-body interactions in the CuZr alloy exhibit strong deviations from additivity, i.e. the repulsive component of the Cu-Zr and Zr-Zr potentials exhibiting very similar characteristic lengths, both substantially larger than that of the Cu-Cu steric length. In addition, we find that the attractive wells in 2-body interactions for the CuZr alloy are significantly shallower and more extended than is found in NiAl.

Based on these 2-body interactions alone, we would expect that the B2 crystal structure would be significantly destabilized in the case of CuZr since this crystal typically requires an $d_{SL}$ length that is $\approx 0.86\ d_{LL}$, where $d_{SL}$ and $d_{LL}$ are the distances between small-large and large-large nearest neighbours, respectively. Since the 3-body contributions in CuZr are all repulsive in character, the absence of significant attraction in the CuZr 2-body interactions implies that the cohesion of this alloy rests on the embedding field contribution given by $F_{Cu}(\bar{\rho}_{Cu})$ and $F_{Zr}(\bar{\rho}_{Zr})$.

The reasoning of the preceding paragraph is based on the dominance of 2-body interactions as expressed in the pairwise additive models. The applicability of this reasoning must be questioned in the case of metallic liquids. To understand how the interaction energy differs between the two alloys, we need to assess the contributions of the one body and three body terms, as well as the pairwise contribution, to the potential energy. In Table 1 we present the values of the total potential energy for liquid and crystal states of the two alloys calculated exactly (i.e. within the model implemented using LAMMPS) and using the $2^{nd}$ order expansion derived above. Also presented are the contributions to the approximate energy for the one, two and three body terms as given in Eqs. 22-29. Considering the data in Table 1, we note that the $2^{nd}$ order expansion of the energy provides a good account of the energy in both alloys and for both states.

The key result of Table 1 is to be found in the differences between the crystal and liquid energies. In both liquids, it is the single body potential, not the two body attractions, that drives crystallization. In fact, the two body energy *increases* on crystallizing, in stark contrast to the experience of freezing of Lennard-Jones liquids. Since, in both alloys, the single body terms drive crystallization by favouring densification and the two body energies are forced to increase to accommodate this change, it appears to be the three body terms that provides the clearest difference between the two alloys. In CuZr, this latter contribution is purely repulsive while in NiAl, negative energies are found for half of the triplets. On freezing of NiAl, these negative triplet contributions produce a substantial energy decrease, roughly half that of the one body contribution. In the freezing of CuZr, the stabilizing contribution is relatively smaller, roughly a third of the one body energy change.

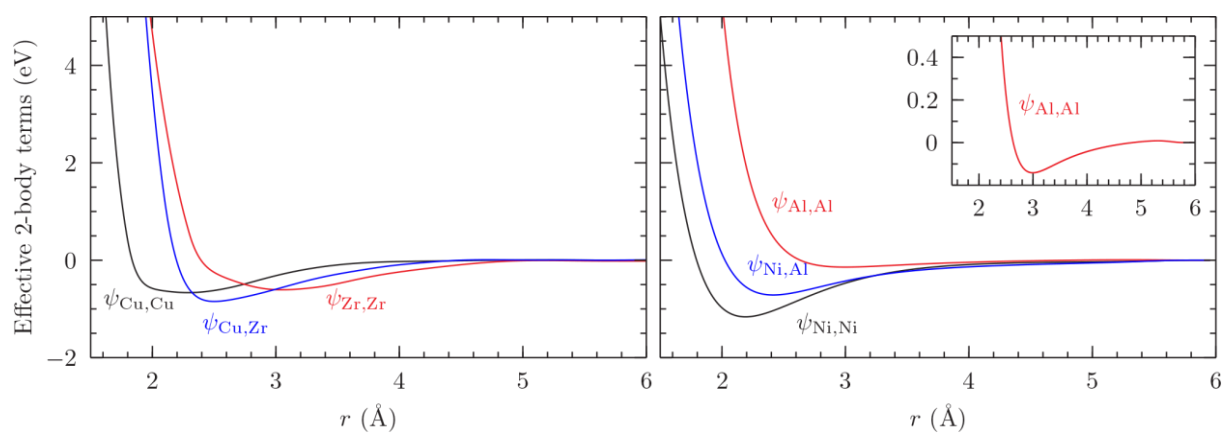

**Figure 7**. The effective 2-body potentials for the alloys $Cu_{50}Zr_{50}$ (left panel) and $Ni_{50}Al_{50}$ (right panel) calculated using Eqs. 23-27. Insert. The attractive tail of the Al-Al interaction viewed over finer resolution of energy.

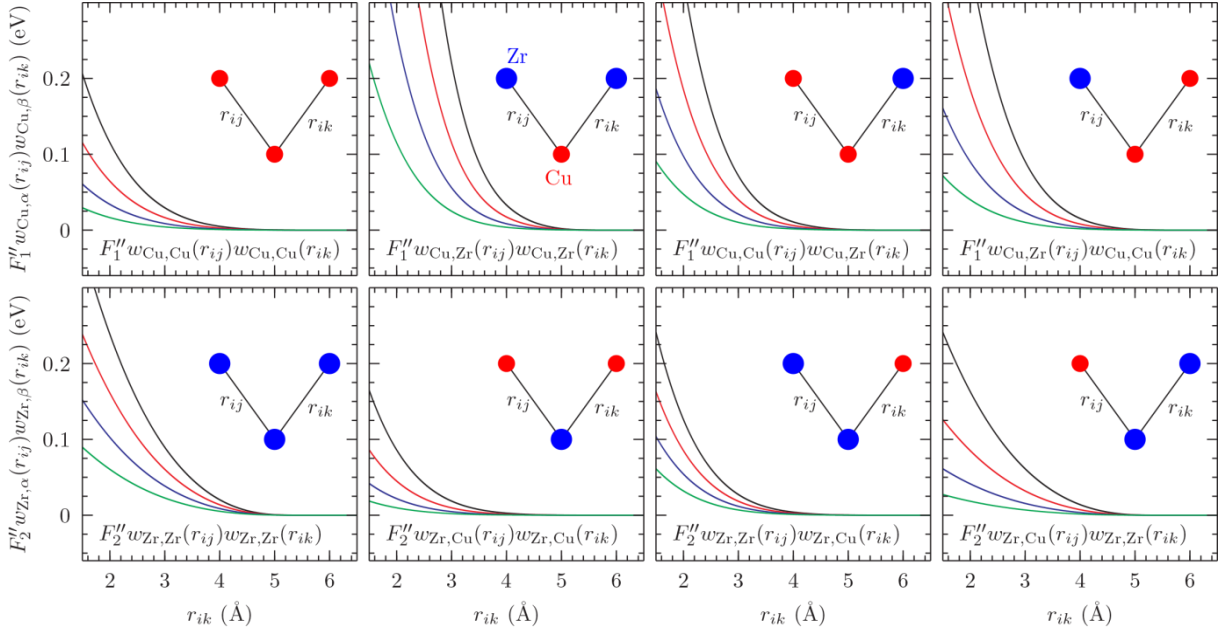

**Figure 8.** The individual contributions to the effective 3-body potentials for $Cu_{50}Zr_{50}$ at some fixed values of $r_{ij}$, i.e. $r_{ij}$ =1.5 Å (black curve), 2 Å (red curve), 2.5 Å (blue curve) and 3 Å (green curve) at T = 1100K. These interactions have been calculated using the terms in Eqs. 29. The included figures indicate the composition arrangement of each 3-atom configuration, with the large blue dot being Zr and small red dot being Cu.

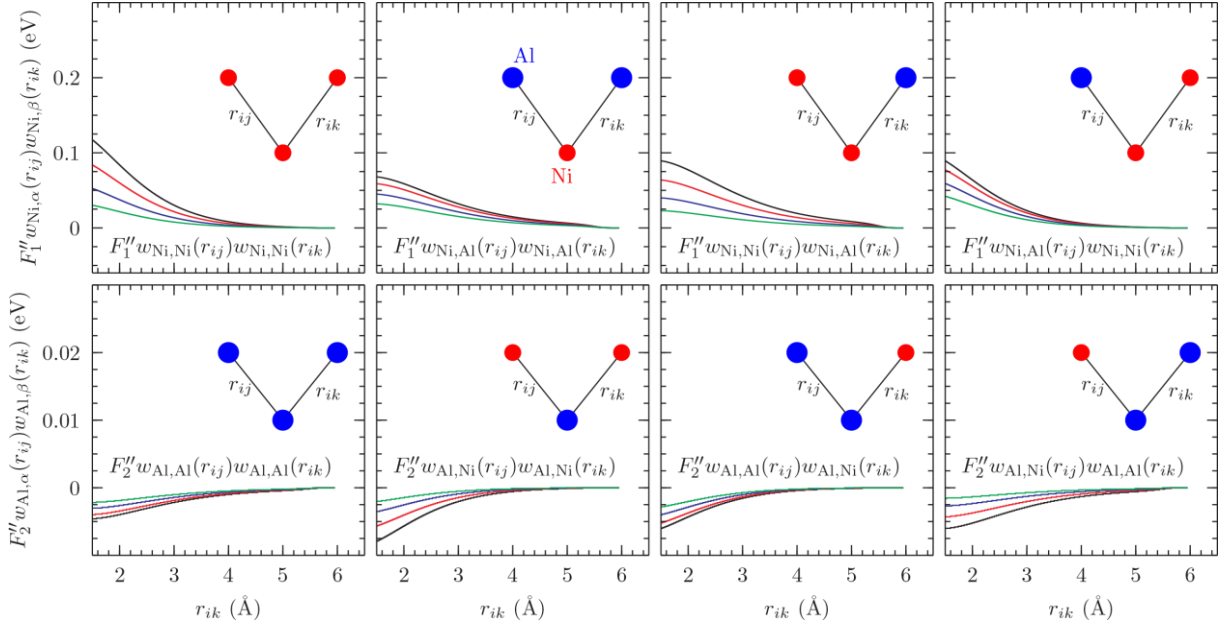

**Figure 9.** The individual contributions to the effective 3-body potentials for $Ni_{50}Al_{50}$ at some fixed values of $r_{ij}$, i.e. $r_{ij}$ =1.5 Å (black curve), 2 Å (red curve), 2.5 Å (blue curve) and 3 Å (green curve) at T = 1300K. These interactions have been calculated using the terms in Eq. 29. The included figures indicate the composition arrangement of each 3-atom configuration, with the large blue dot being Al and small red dot being Ni.

| System | PE | PE$_{approx}$ | One-body | Two-body | Three-body |
|---|---|---|---|---|---|
| Liquid CuZr | -4.716 | -4.725 | -2.087 | -3.715 | 1.076 |
| Crystal CuZr | -4.897 | -4.907 | -2.342 | -3.549 | 0.985 |
| **Δ$_{Cryst-Liq}$** | **-0.181** | **-0.182** | **-0.255** | **+0.166** | **-0.091** |
| Liquid NiAl | -4.056 | -4.052 | -1.260 | -4.340 | 1.548 |
| Crystal NiAl | -4.273 | -4.245 | -1.846 | -3.655 | 1.255 |
| **Δ$_{Cryst-Liq}$** | **-0.217** | **-0.193** | **-0.586** | **+0.685** | **-0.293** |

**Table 1**. The potential energies of liquid/crystal CuZr at 1100 K and liquid/crystal NiAl at 1300 K, obtained from the EAM potentials (PE) and from the approximation based on the Taylor expansion of the potential (PE$_{approx.}$), respectively. The contributions of one-body, two-body, and three-body interactions to PE$_{approx}$ are also listed. The difference between the crystal and liquid terms are presented in the rows labelled Δ$_{Cryst-Liq}$. The unit of energy is eV/atom.

## 6. Conclusion

The conclusions of this paper are as follows. i) The chemical ordering in the liquid NiAl is consistent with that induced by the crystal structure (with its characteristic dominance by the larger particle length scale) while the weaker chemical ordering in CuZr shows less evidence of this connection. ii) The BT correlations functions do not, in the context of these two model alloys, provide any clear benefit over the partial correlation functions in differentiating the structure in the two alloys. iii) We have derived a new property, $\hat{R}_{cn}(q)$, the wave-vector dependent susceptibility of the composition to density fluctuations, and shown that it vanishes in the CuZr alloy at wavelengths associated with crystal ordering, while the analogous susceptibility in NiAl does not. The significance of this result is qualified by the large variation of $\hat{R}_{cn}(q)$ in the range of interest. iv) We have presented a direct comparison of the potentials for the two model alloys, using a $2^{nd}$ order density expansion, to arrive at a unique decomposition of the interactions in each system into one, two and three body contributions. We established that the one body energy plays a crucial role in stabilizing the crystal relative to the liquid in both alloys but that, in the case of the NiAl alloy, the three body contribution to the heat of fusion is also substantial, considerably larger than the analogous contribution in CuZr.

Looking through these results we acknowledge that none of them can provide a clear rationale for the striking difference in crystallization kinetics of these two alloys. The three body contributions to the potential energy provide some clear difference between the two alloys but none that point directly to a difference in glass forming ability. While it is perilous to draw strong conclusions from null outcomes, this study lends some support to the suggestion that the quest for an explanation of crystallization kinetics based solely on liquid properties may be futile. We have already established [19] that free liquid surface can induce composition fluctuation in NiAl but not CuZr. As neither the BT correlations nor our linear response analysis could reproduce this behaviour, we conclude that the surface induced compositional ordering in NiAl represents a *nonlinear* response of the liquid. Crystallization kinetics, in other words, may be most significantly governed by the differing response of liquids to the large perturbations associated with crystallization and not the response accessible in the (metastable) equilibrium properties of the supercooled liquid.

**Acknowledgements**

We acknowledge support from the Australian Research Council. CT would particularly like to thank the ARC for the DECRA Fellowship (Grant No. DE150100738) and the NCI National Facility for computational support of project code eu7.

**Appendix. Derivation the Expression for the Response Function $\hat{R}_{nc}(q)$ in Equation 16.**

We begin with Eq. 9 which establishes the linear relation that describes the density variation $\delta n(\vec{r})$ as a functional of the potential $u_n(\vec{r})$, i.e.

$$\delta n(\vec{r}_1) = \int d\vec{r}_2 \, G_{nn}(\vec{r}_1, \vec{r}_2) u_n(\vec{r}_2) \tag{A1}$$

We can formally invert this relation to provide an expression of the potential $u_n(\vec{r})$ as a functional of $\delta n(\vec{r})$ by introducing the inverse of the susceptibility $G_{nn}(r_1,r_2)$,

$$\int d\vec{r}_1 G_{nn}^{-1}(\vec{r}_3, \vec{r}_1) \delta n(\vec{r}_1) = \int d\vec{r}_2 \int d\vec{r}_1 G_{nn}^{-1}(\vec{r}_3, \vec{r}_1) G_{nn}(\vec{r}_1, \vec{r}_2) u_n(\vec{r}_2) = u_n(\vec{r}_3) \tag{A2}$$

where

$$\int d\vec{r}_1 G_{nn}^{-1}(\vec{r}_3, \vec{r}_1) G_{nn}(\vec{r}_1, \vec{r}_2) = \delta(\vec{r}_3, \vec{r}_2) \tag{A3}$$

with $\delta(\vec{r}_3, \vec{r}_2)$ being the Dirac delta function, defines the functional inverse.

In Eq. 8 we considered $\delta c(\vec{r})$ as a functional of $u_n(\vec{r})$,

$$\delta c(\vec{r}_1) = \int d\vec{r}_2 G_{nc}(\vec{r}_1, \vec{r}_2) u_n(\vec{r}_2) \tag{A4}$$

Now we can replace $u_n(\vec{r})$ in Eq.A4 by $\delta n(\vec{r})$ by substituting Eq. A2 into Eq. A4 to give

$$\delta c(\vec{r}_1) = \int d\vec{r}_2 R_{cn}(\vec{r}_1, \vec{r}_2) \delta n(\vec{r}_2) \tag{A5}$$

where

$$R_{cn}(\vec{r}_1, \vec{r}_2) = \int d\vec{r}_3 G_{nn}^{-1}(\vec{r}_1, \vec{r}_3) G_{nc}(\vec{r}_3, \vec{r}_2) \tag{A6}$$

Finally, we can take the Fourier transform of Eqs. A5 and A6 to get

$$\delta \hat{c}(q) = \hat{R}_{cn}(q) \delta \hat{n}(q) \tag{A7}$$

and

$$\hat{R}_{cn}(q) = \hat{G}_{nn}^{-1}(q) \hat{G}_{nc}(q) \tag{A8}$$

where

$$\hat{G}_{nc}(q) = n_o S_{nc}(q) \tag{A9}$$

and

$$\hat{G}_{nn}^{-1}(q) = \frac{1}{n_o S_{nn}(q)} \tag{A10}$$

So, finally, the desired response function $\hat{R}_{cn}(q)$ (i.e. Eq. 16) can be calculated as

$$\hat{R}_{cn}(q) = \frac{S_{nc}(q)}{S_{nn}(q)} \tag{A11}$$